\documentclass[conference]{IEEEtran}
\IEEEoverridecommandlockouts
\usepackage{cite}
\usepackage{amsmath,amssymb,amsfonts}
\usepackage{epsfig}
\usepackage{theorem}
\usepackage{algorithmic}
\usepackage{graphicx}
\usepackage{textcomp}
\usepackage{xcolor}
\def\BibTeX{{\rm B\kern-.05em{\sc i\kern-.025em b}\kern-.08em
    T\kern-.1667em\lower.7ex\hbox{E}\kern-.125emX}}

\DeclareMathOperator{\sinc}{sinc} 
\DeclareMathOperator{\rect}{rect}
\DeclareMathOperator{\atan}{atan}

\begin{document}

\title{Characterizing the Ambiguity Function of Constant-Envelope OFDM Waveforms
\thanks{David G. Felton's efforts were supported by the Naval Research Enterprise Internship Program (NREIP) and David A. Hague's efforts were supported by the Naval Undersea Warfare Center's In-House Laboratory Independent Research (ILIR) program.}
}

\author{\IEEEauthorblockN{David G. Felton$^1$ David A. Hague$^2$}
\IEEEauthorblockA{\textit{$^1$Radar Systems Lab (RSL), University of Kansas, Lawrence, KS}\\
\textit{$^2$Sensors and Sonar Systems Deptartment, Naval Undersea Warfare Center, Newport, RI}}
}
\maketitle

\begin{abstract}
This paper investigates the radar Ambiguity Function (AF) properties of Constant Envelope Orthogonal Frequency Division Multiplexing (CE-OFDM) waveforms employing Phase-Shift Keying (PSK).  The CE-OFDM is in fact a special case of the Multi-Tone Sinusoidal Frequency Modulated (MTSFM) waveform which allows for applying many of the same mathematical techniques of the MTSFM model to the CE-OFDM model. This results in novel compact closed-form expressions for the spectrum, AF, and Auto-Correlation Function (ACF) of the CE-OFDM waveform.  The mainlobe structure of the CE-OFDM's AF is characterized by the Ellipse of Ambigiuity (EOA) model.  This produces precise closed-form expressions for the CE-OFDM's Root-Mean Square (RMS) bandwidth and the degree of range-Doppler coupling present in the waveform's AF mainlobe.  These expressions show that a CE-OFDM waveform employing PSK as the symbol encoding scheme will possess a fixed RMS bandwidth for fixed modulation index $h$ and number of sub-carriers $L$.  Additionally, we show that the EOA model predicts that a CE-OFDM waveform employing PSK encoding will almost always possess a ``Thumbtack-Like'' AF shape.  

\end{abstract}


\section{Introduction}
The Constant-Envelope Orthogonal Frequency Division Multiplexing (CE-OFDM) waveform is a constant envelope analogue of the standard OFDM waveform model where OFDM modulation is performed in either the instantaneous phase or frequency domain \cite{Thompson_2, Chung_Comm}.  The CE-OFDM waveform encoding is essentially a form of Frequency Modulation (FM) which guarantees a constant envelope.  This constant envelope property makes these waveforms better suited for transmission on real-world radar and communications transmitters than standard OFDM waveforms \cite{Thompson_1} whose complex envelope can vary substantially \cite{Levanon}.  CE-OFDM waveforms have recently been proposed for use in Dual-Function Radar/Communication (DFRC) applications \cite{Aubry_2016_Optimization, IEEE_Sig_Proc_RadarComm, Zhang_Radar_Comms} as the encoded communication symbols can also be utilized as a discrete set of parameters that realize waveforms with desirable Ambiguity Function (AF) properties \cite{Blunt_CEOFDM}.  

The majority of the literature on CE-OFDM waveforms initially focused on its properties and performance for use as a communications waveform \cite{Chung_Comm, Thompson_1, Thompson_2, Thompson_4, Liu}.  More recent efforts in the literature have studied CE-OFDM as a potential radar waveform and have begun to characterize its properties from the radar waveform design perspective\cite{Mohseni, Blunt_CEOFDM}.  These efforts have mainly focused on analyzing the structure of the waveform's spectrum, AF, and Auto/Cross-Correlation Functions (ACF/CCF).  These first-principle design metrics fundamentally determine a radar waveform's performance.  However, to the best of the authors' knowledge, the properties of the CE-OFDM waveform's AF has not been fully characterized.   Doing so might provide insight into the appropriate selection of symbol encoding for DFRC applications.

This paper takes a first step in more fully characterizing the structure of the CE-OFDM waveform's AF.  We utilize a CE-OFDM waveform model that employs Phase-Shift Keying (PSK) to encode symbols.  We first show that the CE-OFDM waveform is a special case of the Multi-Tone Sinusoidal FM (MTSFM) waveform, an adaptive transmit waveform model from the active sonar literature \cite{Hague_AES, Hague_Comb}.  Leveraging results from the efforts in \cite{Hague_AES}, we derive exact closed-form expressions for the CE-OFDM waveform's spectrum, AF, and ACF which to the best of the authors' knowledge appear to be novel.  We additionally fully characterize the mainlobe structure of the CE-OFDM waveform's AF using the Ellipse of Ambiguity (EOA) model from the well established radar literature \cite{Cook, Rihaczek}.  We show that a CE-OFDM waveform employing PSK essentially always realizes a ``Thumbtack-Like'' AF shape.  These results also provide an exact closed-form expression for the Root Mean Square (RMS) bandwidth that unlike previous approximations \cite{Thompson_1} is free from any restrictions on the choice of CE-OFDM waveform parameters.  Finally, we show that the metrics describing the CE-OFDM's ACF sidelobe structure are multi-modal functions of the waveform's PSK symbols. Thus, the search for ``optimal'' codes that produce desirably low AF/ACF sidelobes will require either a completely different code construction scheme from that of Phase-Coded waveforms \cite{Levanon} or direct numerical optimization methods.  

\section{The Structure of the Ambiguity Function}
\label{sec:AF}
In general, a basebanded FM waveform is expressed in continuous time as
\begin{equation}
s\left(t\right)=\frac{\rect\left(t/T\right)}{\sqrt{T}} e^{j\varphi\left(t\right)}
\label{eq:complexExpo}
\end{equation}
where $T$ is the waveform's duration and $\varphi\left(t\right)$ is its phase modulation function. The time support is $|t| \leq T/2$, and $1/\sqrt{T}$ normalizes the signal energy to unity.  The waveform's frequency modulation function $m\left(t\right)$ maps its instantaneous frequency as a function of time and is expressed as
\begin{equation}
m\left(t\right) = \dfrac{1}{2\pi}\dfrac{\partial\varphi(t)}{\partial t}.
\label{eq:modFunc}
\end{equation}
The AF measures the response of the waveform's MF to its Doppler shifted versions and is defined as \cite{Rihaczek, Cook}
\begin{IEEEeqnarray}{rCl}
\chi\left(\tau, \nu\right) = \int_{-\infty}^{\infty}s\left(t-\frac{\tau}{2}\right)s^*\left(t+\frac{\tau}{2}\right)e^{j2\pi \nu t} dt
\label{eq:AF}
\end{IEEEeqnarray}
where $\nu$ is the Doppler shift.  Lastly, the ACF is the zero Doppler cut of the AF
\begin{IEEEeqnarray}{rCl}
R\left(\tau\right) = \chi\left(\tau, 0\right) = \int_{-\infty}^{\infty}s\left(t-\frac{\tau}{2}\right)s^*\left(t+\frac{\tau}{2}\right) dt.
\label{eq:ACF}
\end{IEEEeqnarray}
There are a series of design metrics that characterize a waveform's AF shape in terms of its mainlobe and sidelobe structure.  These metrics are described below.

\subsubsection{Mainlobe Structure}
\label{subsubsec:mainlobe}
The AF mainlobe structure determines a waveform's ability to estimate the range and Doppler of a target and to resolve multiple targets in range and Doppler.  The AF mainlobe can be approximated by a second order Taylor series expansion \cite{Cook, Rihaczek}.  The EOA is the contour of this AF mainlobe approximation at some height $\xi$ and is always a coupled ellipse \cite{Cook}.  The EOA for the AF is expressed as \cite{Cook, Rihaczek, Ricker}
\begin{equation}
1 - |\chi\left(\tau, \nu\right)|^2 = \xi = \beta_{rms}^2\tau^2 + 2\rho\tau\nu + \tau_{rms}^2\nu^2
\end{equation}
where $\beta_{rms}^2$ is the waveform's RMS bandwidth and determines time-delay (range) sensitivity, $\tau_{rms}^2$ is the RMS pulse-length which determines Doppler (range-rate) sensitivity, and $\rho$ is the Range-Doppler Coupling Factor (RDCF) for the AF mainlobe.  

The RMS bandwidth is expressed as \cite{Ricker}
\begin{align}
\beta_{rms}^2 &= \int_{\infty}^{\infty}\left(f - f_0\right)^2 |S\left(f\right)|^2 df \\ &= \int_{\Omega_t} | \dot{s}\left(t\right)|^2 dt - \left| \int_{\Omega_t} s\left(t \right) \dot{s}^*\left(t\right) dt \right| ^2 \\ &= \dfrac{1}{T}\int_{-T/2}^{T/2} \left[\dot{\varphi}\left( t \right)\right]^2 dt - \left| \dfrac{1}{T}\int_{-T/2}^{T/2} j \dot{\varphi}\left(t\right) dt \right| ^2
\label{eq:Brms}
\end{align}
where $f_0$ is the waveform's spectral centroid $\langle f \rangle$, $S\left(f\right)$ is the waveform's Fourier transform, $\dot{s}\left(t\right)$ is the first time derivative of the waveform $s\left(t\right)$, $\Omega_t$ represents the region of support in time of the waveform, and $\dot{\varphi}\left(t\right)=2\pi m\left(t\right)$ is the first time derivative of the waveform's instantaneous phase.  Note that \eqref{eq:Brms} results from inserting \eqref{eq:complexExpo} into (7).  The RMS pulse-length term is expressed as 
\begin{equation}
\tau_{rms}^2 = 4\pi^2\int_{\Omega_t} \left(t-t_0\right)^2 |s\left(t\right)|^2 dt
\label{eq:Trms}
\end{equation}
where $t_0$ is the first time moment $\langle t \rangle$ of the the waveform $s\left(t\right)$ and is zero for waveforms such as \eqref{eq:complexExpo} that are even-symmetric in time.  The RDCF is expressed as 
\begin{equation}
\rho = -2\pi \Im \Biggl\{\int_{\Omega_t} ts\left(t\right)\dot{s}^*\left(t\right) dt \Biggr\} = 2\pi  \int_{-T/2}^{T/2}  t \dot{\varphi}\left(t\right) dt.
\label{eq:rho}
\end{equation}
where $\Im \{\}$ denotes the imaginary component of the integral.  The second expression in \eqref{eq:rho} results from inserting \eqref{eq:complexExpo} into the first expression in \eqref{eq:rho}.  

For the waveform model in \eqref{eq:complexExpo}, the RMS pulse-length $\tau_{rms}^2$ is solely dependent upon the pulse-length $T$ of the waveform.  The RMS bandwidth and RDCF EOA parameters are solely dependent upon the modulation function \eqref{eq:modFunc}.  If the waveform's modulation function is known, $\beta_{rms}^2$ and $\rho$ can be calculated in exact closed form which along with $\tau_{rms}^2$ provides full characterization of the waveform's AF mainlobe structure \cite{Cook, Rihaczek, Ricker}.  Furthermore, if the waveform's modulation function is composed of a discrete set of parameters, those parameters can be chosen to design waveforms with specific EOA parameters thus directly shaping the AF mainlobe structure \cite{Hague_SSP_2018}.

\subsubsection{Sidelobe Structure}
\label{subsubsec:sidelobe}
This paper focuses specifically on the sidelobe structure of the ACF.  Two of the most common metrics are the Peak-to-Sidelobe Level Ratio (PSLR) and the Integrated Sidelobe Level (ISL).  The PSLR is expressed as
\begin{IEEEeqnarray}{rCl}
\text{PSLR} =  \dfrac{\underset{\Delta\tau \leq \vert\tau\vert \leq T}{\text{max}}\bigl\{\left|R\left(\tau\right)\right|^2\bigr\}}{\underset{0 \leq \vert\tau\vert \leq \Delta \tau}{\text{max}}\bigl\{\left|R\left(\tau\right)\right|^2\bigr\}} = \underset{\Delta \tau \leq \vert\tau\vert \leq T}{\text{max}}\bigl\{\left|R\left(\tau\right)\right|^2\bigr\}.
\label{eq:PSLR}
\end{IEEEeqnarray} 
where $\Delta \tau$ is the null of the ACF mainlobe.  Thus, the ACF's null-to-null mainlobe width is $2\Delta \tau$.  Note that the rightmost expression in \eqref{eq:PSLR} results from the assumption that the waveform is unit energy and thus the maximum value of $|R\left(\tau\right)|^2$ is unity which occurs at $\tau = 0$.  The ISL is the ratio of the area under the sidelobe region $A_{\tau}$ of $|R\left(\tau\right)|^2$ to the area under mainlobe region $A_0$ of $|R\left(\tau\right)|^2$ expressed as
\begin{IEEEeqnarray}{rCl}
\text{ISL}~=\dfrac{A_{\tau}}{A_0} = \dfrac{\int_{\Delta \tau}^{T}\left|R\left(\tau\right)\right|^2 d\tau}{\int_{0}^{\Delta \tau}\left|R\left(\tau\right)\right|^2 d\tau}.
\label{eq:ISL}
\end{IEEEeqnarray}
Note that the integration is performed only over positive time-delays since the ACF is even-symmetric in $\tau$.  A lower ISL  corresponds to an ACF with lower overall sidelobe levels but does not necessarily translate to a lower PSLR.

\section{CE-OFDM Waveform Model}
\label{sec:CE-OFDM}
This section describes how the CE-OFDM waveform model is derived from the standard OFDM waveform.  Additionally, this section shows how the CE-OFDM waveform model is closely related to the MTSFM waveform model \cite{Hague_AES} and shares many of its mathematical properties.  

\subsection{The CE-OFDM Waveform}
\label{subsec:CE-OFDM}
The CE-OFDM waveform is realized as a phase-modulated waveform whose instantaneous phase itself takes on the form of an OFDM waveform
\begin{equation}
\varphi\left(t\right) = 2\pi h\sum_{\ell=-L}^L c_{\ell} e^{j\frac{2\pi \ell t}{T}}
\label{eq:ceofdm_1}
\end{equation}
where $c_{\ell}$ are the complex valued PSK symbols, $L$ is the number of unique complex sub-carriers, and $h$ is the waveform's modulation index which in conjunction with $L$ controls the bandwidth of the waveform.  Converting the complex Fourier series in \eqref{eq:ceofdm_1} to the real-valued Fourier series results in the modified expression for the CE-OFDM's phase modulation function
\begin{equation}
\varphi\left(t\right) = \dfrac{\tilde{\alpha}_0}{2}+2\pi h \sum_{\ell=1}^L \tilde{\alpha}_{\ell}\cos\left(\dfrac{2\pi \ell t}{T}\right)+\tilde{\beta}_{\ell}\sin\left(\dfrac{2\pi \ell t}{T}\right)
\label{eq:ceofdm_2}
\end{equation}
where $\tilde{\alpha}_{\ell}$ and $\tilde{\beta}_{\ell}$ are the real Fourier series coefficients expressed as
\begin{equation}
\tilde{\alpha}_0 = c_0,~\tilde{\alpha}_{\ell} = \dfrac{c_{\ell} + c_{-\ell}}{2},~\tilde{\beta}_{\ell} = \dfrac{j\left(c_{\ell} - c_{-\ell}\right)}{2}.
\end{equation}

Assuming PSK symbols are used in the CE-OFDM model, the real Fourier series coefficients lie on the unit circle such that $\sqrt{\tilde{\alpha}_{\ell}^2 + \tilde{\beta}_{\ell}^2} = 1$.  Lastly, assuming $\tilde{\alpha}_0=0$ and expressing \eqref{eq:ceofdm_2} in terms of the amplitude-phase representation for real-valued Fourier series results in the final form of the CE-OFDM's phase modulation function
\begin{equation} 
\varphi\left(t\right)=2\pi h\sum_{\ell=1}^L{ \vert\Gamma_{\ell}\vert\cos\left(\dfrac{2\pi \ell t}{T}+\phi_{\ell}\right)}
\label{eq:CE-OFDM}
\end{equation}
where $\vert\Gamma_{\ell}\vert = \sqrt{\tilde{\alpha}_{\ell}^2 + \tilde{\beta}_{\ell}^2} = 1$ and the symbol phases $\phi_{\ell}$ are expressed as 
\begin{equation}
\phi_{\ell} = \atan\left\{\dfrac{\tilde{\beta}_{\ell}}{\tilde{\alpha}_{\ell}} \right\}.
\end{equation}
Using \eqref{eq:modFunc}, the CE-OFDM waveform's frequency modulation function is expressed as
\begin{equation}
m(t)=-\dfrac{2\pi h}{T}\sum_{{\ell}=1}^L\ell\vert\Gamma_{\ell}\vert\sin\left(\dfrac{2\pi \ell t}{T}+\phi_{\ell}\right).
 \label{eq:freq}
\end{equation}
We wish to note here that the CE-OFDM's instantenous phase \eqref{eq:CE-OFDM}, which has the equivalent representation given by \eqref{eq:ceofdm_2}, is in fact a special case of the MTSFM waveform model described in \cite{Hague_AES}.  The MTSFM's instantaneous phase is a more general version of \eqref{eq:ceofdm_2} where $\tilde{\alpha}_{\ell}$ and $\tilde{\beta}_{\ell}$ are allowed to vary in magnitude rather than strictly lie on the unit circle.  Therefore, many of the same mathematical methods used to describe the MTSFM waveform are directly applicable to the CE-OFDM waveform model. 

\subsection{Mathematical Properties of the CE-OFDM Waveform's AF}
This section describes a series of mathematical properties that characterize the CE-OFDM waveform's AF shape.

\subsubsection{Spectrum, AF, and ACF of the CE-OFDM Waveform}
\label{subsubsec:GBFMadness}
Much like the MTSFM waveform model \cite{Hague_AES}, the CE-OFDM waveform time series can be expressed in terms of a complex Fourier series.  As shown in Appendix \ref{sec:AppendixI}, exploiting the Jacobi-Anger expansion for Generalized Bessel Functions (GBFs) \cite{DattoliII, Lorenzutta} results in the following expression for the CE-OFDM waveform's time-series 
\begin{multline}
s\left(t\right)=\frac{\rect\left(t/T\right)}{\sqrt{T}} \times \\ \sum_{m=-\infty}^{\infty}j^me^{jm\phi_1} \mathcal{J}_m\left(\left\{z_{\ell}\right\}_{\ell=1}^L;\left\{\gamma_{\ell}\right\}_{\ell=2}^L\right) e^{j\frac{2\pi m t}{T}}
 \label{eq:sGBF}
\end{multline}
where $\mathcal{J}_m\left(\left\{z_{\ell}\right\}_{\ell=1}^L;\left\{\gamma_{\ell}\right\}_{\ell=2}^L\right)$ is the $L$-dimensional, $(L-1)$-parameter GBF with $L$-dimensional variables $\left\{z_{\ell}\right\}_{\ell=1}^L=2\pi h |\Gamma_{\ell}|$ and $(L-1)$ parameters $\left\{\gamma_{\ell}\right\}_{\ell=2}^L= j^{-(\ell-1)}e^{-j\ell\phi_1}e^{j\phi_{\ell}}$.  Throughout the rest of this paper, the GBF in \eqref{eq:sGBF} will be denoted as simply $\mathcal{J}_m$ for brevity.  Equation \eqref{eq:sGBF} is a more compact representation of the CE-OFDM in time than previous efforts such as those in \cite{Blunt_CEOFDM} which utilize product-of-sums of 1-dimensional Bessel functions.  More importantly, \eqref{eq:sGBF} allows for deriving compact closed-form expressions of the CE-OFDM waveform's spectrum, AF, and ACF.  These expressions, derived in Appendix \ref{sec:AppendixII}, are shown below
\begin{equation}
S\left(f\right) = \sqrt{T}\sum_{m=-\infty}^{\infty}j^me^{jm\phi_1} \mathcal{J}_m \sinc\left[\pi T\left(f-\frac{m}{T}\right)\right],
\label{eq:spectrumGBF}
\end{equation}
\begin{multline} 
\chi\left(\tau,\nu\right)=\left(\frac{T-|\tau|}{T}\right) \sum_{m,n} j^{(m-n)}e^{j\phi_1(m-n)} \mathcal{J}_m \mathcal{J}_n^* \\ \times e^{-j\frac{\pi\left(m+n\right)\tau}{T}} \sinc\left[\pi \left(\frac{T-|\tau|}{T}\right)\left(\nu T +\left(m-n\right)\right) \right].
\label{eq:AFgbf}
\end{multline}
The ACF of the CE-OFDM follows from setting $\nu$ in \eqref{eq:AFgbf} to zero.  The expression for the CE-OFDM's spectrum \eqref{eq:spectrumGBF} and AF \eqref{eq:AFgbf} are closely related to that of the MTSFM's spectrum and AF derived in \cite{Hague_AES} with the only difference being the type of GBF in the expression and the complex phase terms.  

\subsubsection{Mainlobe Characterization}
The EOA parameters of the CE-OFDM, derived in Appendix \ref{sec:AppendixIII}, are expressed as
\begin{align}
\beta_{rms}^2 &= \dfrac{4\pi^4h^2}{3T^2}\left(2L^3+3L^2+L\right), \label{eq:Brms1}\\
\tau_{rms}^2 &= \dfrac{\pi^2T^2}{3}, \label{eq:Trms1}\\
\rho &= 4\pi^2h\sum_{\ell=1}^L\vert \Gamma_{\ell} \vert \left(-1\right)^{\ell}\cos\left(\phi_{\ell}\right). \label{eq:rho1}
\end{align}
We'll denote the RDCF $\rho$ in terms of the normalized RDCF $\tilde{\rho}$ which takes on the values $-1 \leq \rho \leq 1$ \cite{Cook, Rihaczek, Cohen} and is expressed as
\begin{equation}
\tilde{\rho} = \dfrac{\rho}{\tau_{rms}\beta_{rms}} = \left(\dfrac{6}{\pi}\right)\dfrac{\sum_{\ell=1}^L\vert \Gamma_{\ell} \vert \left(-1\right)^{\ell}\cos\left(\phi_{\ell}\right)}{\sqrt{2L^3+3L^2+L}}.
\label{eq:rho2}
\end{equation}
We note several observations regarding the results shown in \eqref{eq:Brms1}-\eqref{eq:rho2}.  First, as with all rectangularly tapered waveforms, the Doppler sensitivity is proportional to the square of the waveform duration $T$ times a constant ($\pi^2/3$).  Second, the expression for $\beta_{rms}^2$ in \eqref{eq:Brms1} is an exact closed-form expression for all real values of $h$ and $L$.  This result strongly contrasts with the approximation to $\beta_{rms}^2$ commonly encountered in the literature of $\beta_{rms}^2 \approx 4\pi^2h^2 L^2/T^2$ which holds for $2\pi h > 1$ \cite{Thompson_1, Thompson_2}.  Additionally, equation \eqref{eq:Brms1} shows that the choice of PSK sequence $\phi_{\ell}$ does not influence $\beta_{rms}^2$.    Lastly, the normalized RDCF $\tilde{\rho}$ in \eqref{eq:rho2} is dependent only on the number of carriers $L$ and the PSK sequence $\phi_{\ell}$ and not on modulation index $h$ or waveform duration $T$.  

\section{Implications for CE-OFDM Waveform Design}
\label{sec:designExamples}

\begin{figure}[ht]
\centering
\includegraphics[width=0.5\textwidth]{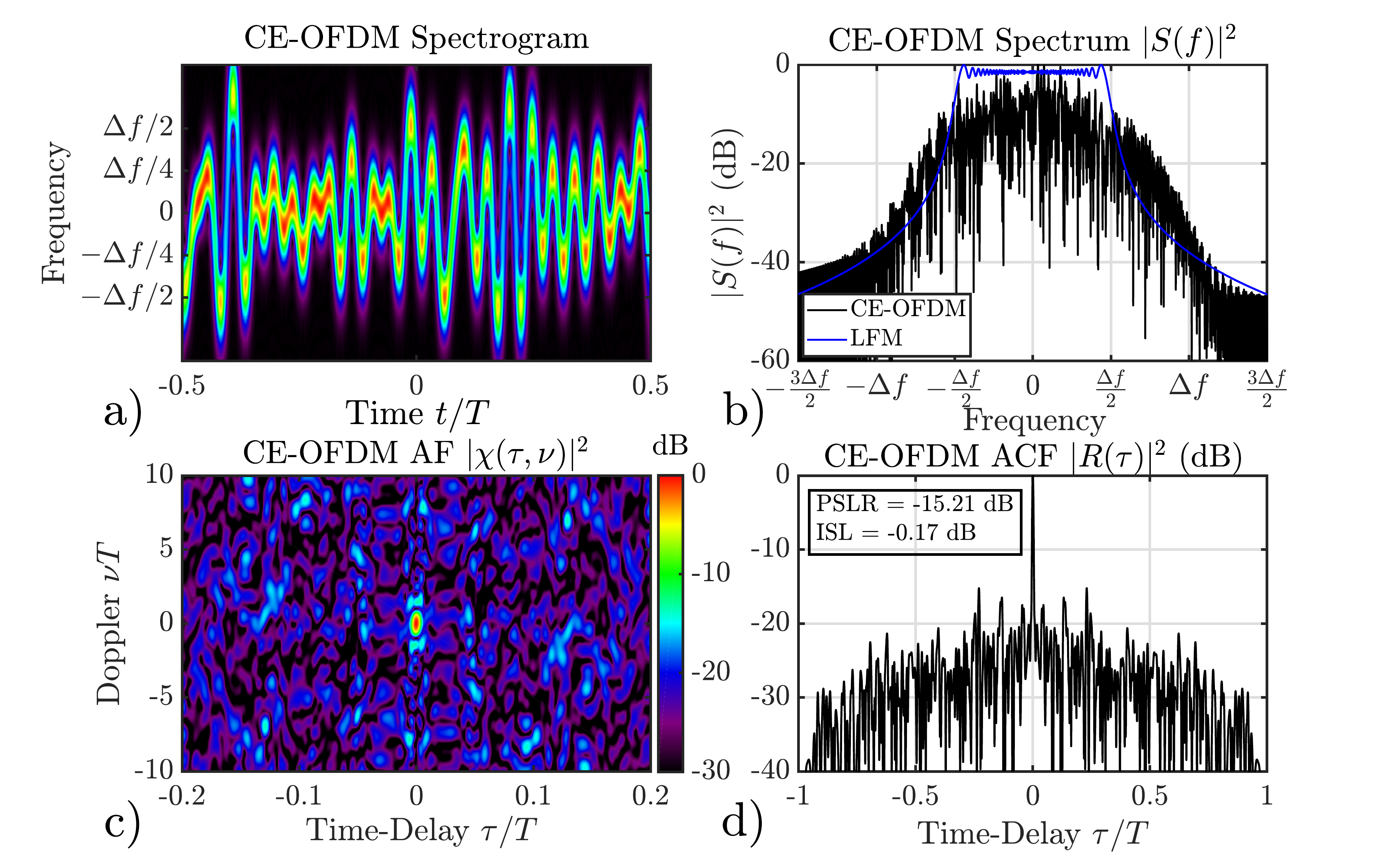}
\caption{Spectrogram (a), spectrum (b), AF (c), and ACF (d) of a CE-OFDM waveform with a TBP of 200 ($h=0.1856$) and $L=24$ randomly generated PSK symbols where $M_{\text{PSK}}=32$.  The CE-OFDM waveform possesses a spectrum that is densely concentrated in a compact band of frequencies and possesses a ``Thumbtack-Like'' AF shape with $\tilde{\rho} = 0.0848$.}
\label{fig:CE-OFDM_1}
\end{figure}

Figure \ref{fig:CE-OFDM_1} shows an example CE-OFDM waveform with $L=24$ sub-carriers with $M_{\text{PSK}} = 32$ symbols $\phi_{\ell}$ drawn randomly.  The waveform's modulation index $h=0.1856$ and is chosen such that its RMS bandwidth \eqref{eq:Brms1} is equivalent to that of an LFM waveform with a Time-Bandwidth Product (TBP) of 200. This is done by setting \eqref{eq:Brms1} equal to the LFM's RMS bandwidth $\pi^2\Delta f^2/3$ and solving for $h$
\begin{align}
\IEEEnonumber \dfrac{4\pi^4h^2}{3T^2}\left(2L^3+3L^2+L\right) &= \dfrac{\pi^2 \Delta f^2}{3} \\ h &= \frac{T\Delta f}{2\pi \sqrt{2L^3+3L^2+L}}
\end{align}
where $\Delta f$ is the LFM waveform's swept bandwidth.  Note that while both waveforms possess the same RMS bandwidth, the CE-OFDM sweeps through a band that is slightly larger than $\pm \Delta f/2$.  However, as can be seen in panel (b) of Figure \ref{fig:CE-OFDM_1}, both waveforms possess spectra whose magnitude falls off at about the same rate outside the $\pm \Delta f / 2$ band.  The waveform possesses a "Thumbtack-Like" AF shape with a largely uncoupled mainlobe ($\tilde{\rho} = 0.0848$) and the ACF of the waveform attains a PSLR of -15.21 dB and an ISL of -0.17 dB.   


\begin{figure}[ht]
\centering
\includegraphics[width=0.5\textwidth]{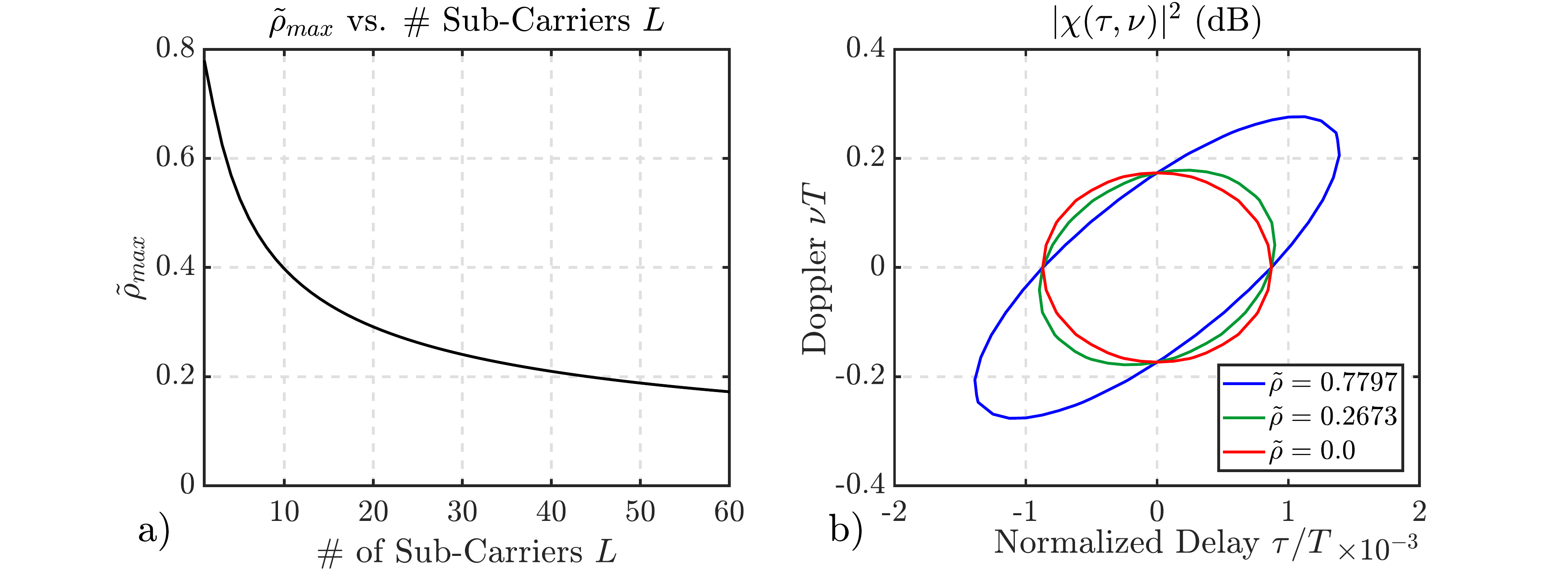}
\caption{Maximum normalized RDCF $\tilde{\rho}_{\text{max}}$ (a) for CE-OFDM waveforms as a function of number of carriers $L$ and several example mainlobe ellipses (b) for varying values of $\tilde{\rho}$.  For practical values of $L$, the CE-OFDM almost always possesses a "Thumbtack-Like" AF shape.}
\label{fig:CE-OFDM_2}
\end{figure}

The ``Thumbtack-like'' AF shape is essentially the only AF shape attainable for a CE-OFDM waveform utilizing PSK.  The max value for the RDCF as a function of $L$, derived in Appendix \ref{sec:AppendixIII}, is expressed as
\begin{equation}
\tilde{\rho}_{\text{max}} = \left(\dfrac{6}{\pi}\right)\dfrac{L}{\sqrt{2L^3+3L^2+L}}
\label{eq:rhoMax}
\end{equation}
and is achieved when the PSK sequence is
\begin{equation}  \phi_{\ell}= \left\{
\begin{array}{ll}
      \pi, & \ell~\text{odd} \\ \\
      
      0, &  \ell~\text{even.} \\
\end{array} 
\right.
\label{eq:rhoMaxSequence}
\end{equation}
Figure \ref{fig:CE-OFDM_2} plots equation \eqref{eq:rhoMax} and several mainlobe ellipses of CE-OFDM waveforms with a TBP of 200 and varying $\tilde{\rho}$ for $\xi = 0.9$.  Note that for increasing $L$, $\tilde{\rho}_{\text{max}}$ falls off at a rate of roughly $L^{-1/2}$ which is for the specific set of PSK symbols $\phi_{\ell}$ shown in \eqref{eq:rhoMaxSequence}.  As a design example, consider the case where $L=24$ like the waveform in Figure \ref{fig:CE-OFDM_1} resulting in $\tilde{\rho}_{\text{max}} = 0.2673$.  As can be seen in Figure \ref{fig:CE-OFDM_2}, this AF mainlobe shape does not substantially differ from a mainlobe with exactly zero coupling.  Even the maximum $\tilde{\rho}_{\text{max}}$ of 0.7797, achieved when $L=1$, produces only a moderately coupled AF mainlobe.  In general, CE-OFDM waveforms utilizing pseudo-random PSK codes tend to produce smaller $\tilde{\rho}$ values than $\tilde{\rho}_{\text{max}}$.  For most CE-OFDM waveform design examples in the literature \cite{Blunt_CEOFDM, Liu, Mohseni}, $L > 24$.  Therefore, CE-OFDM waveforms using PSK encoding and practical values of $L$ will essentially always possess a "Thumbtack-Like" AF shape.  

\begin{figure}[ht]
\centering
\includegraphics[width=0.5\textwidth]{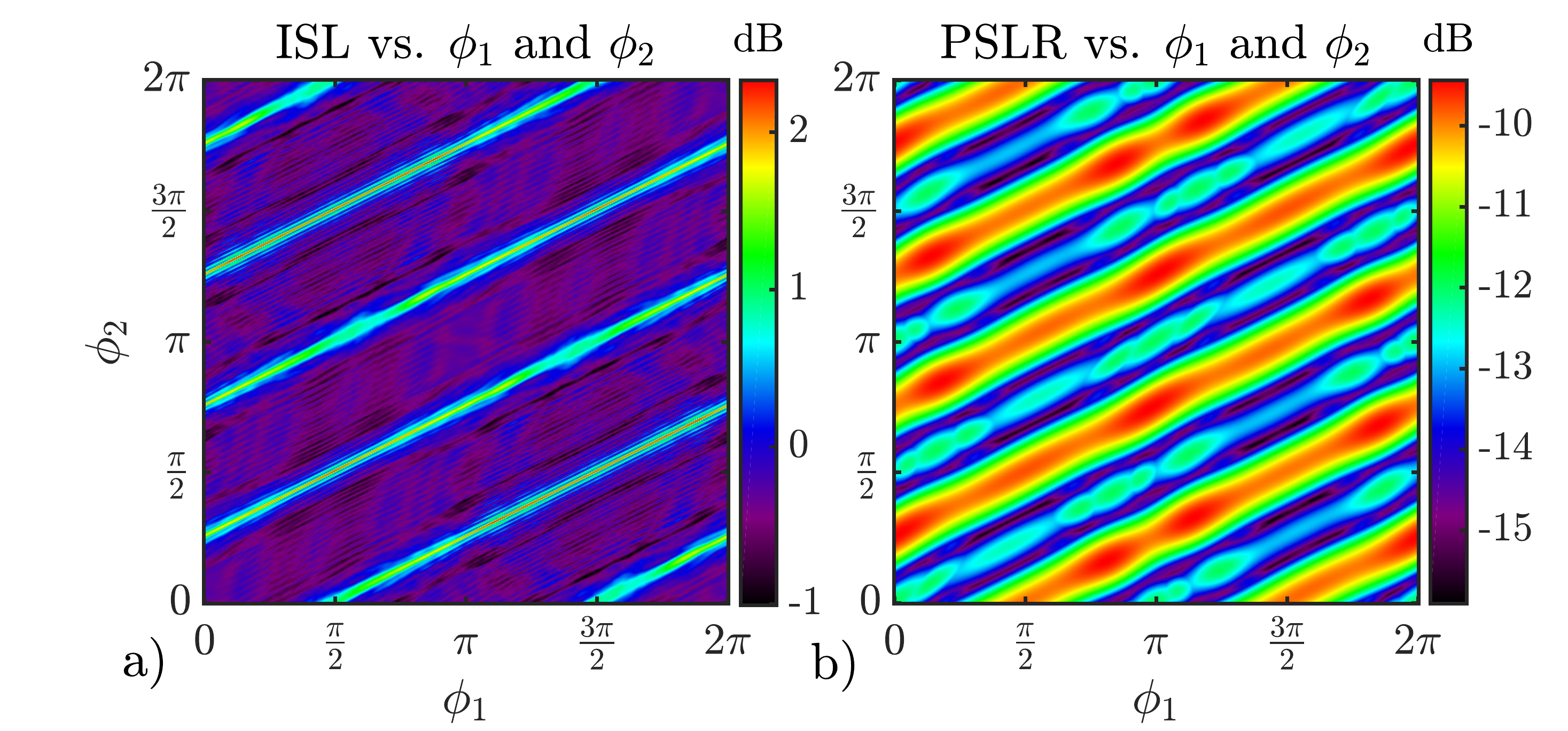}
\caption{Plot of ISL and PSLR vs. $\phi_{\ell}$ for a CE-OFDM waveform with equivalent TBP of 200 and $L=2$ sub-carriers.  Both sidelobe metrics exhibit odd-symmetry in $\phi_1$ and $\phi_2$ as well as multiple local extrema showing that the choice of PSK symbols $\phi_{\ell}$ directly impacts the CE-OFDM waveform's ACF sidelobe structure.}
\label{fig:CE-OFDM_3}
\end{figure}

Figure \ref{fig:CE-OFDM_3} shows the ISL and PSLR of an example CE-OFDM waveform with an equivalent TBP of 200 like the example shown in Figure \ref{fig:CE-OFDM_1} but with only $L=2$ sub-carriers and $h = 5.81$.  Both sidelobe metrics exhibit multiple local extrema and odd-symmetry in $\phi_1$ and $\phi_2$.  This shows that appropriate choice of $\phi_{\ell}$ can indeed influence the CE-OFDM's AF/ACF sidelobe structure.  The symmetry in these metrics suggest that there may exist certain PSK codes that produce more favorably low sidelobe levels.  An alternative to deriving specific codes is to develop waveform optimization methods that produce CE-OFDM waveforms with desirably low ACF sidelobes.  This is the topic of another paper \cite{Hague_Felton_CE_OFDM_2}.

\section{Conclusion}
\label{sec:Conclusion}
This paper derived several exact closed form expressions that more fully characterize the AF shape of CE-OFDM waveforms employing PSK encoding.  These results show that such CE-OFDM waveforms almost always possess a "Thumbtack-Like" AF.  Additionally, we show that the choice of PSK code indeed influences the waveform's ACF sidelobe structure.  Designing CE-OFDM waveforms with desirably low AF/ACF sidelobes will likely involve novel code construction techniques or waveform optimization methods.  Future efforts will explore these code construction and waveform optimization methods as well as extend the analysis performed in this paper to other forms of digital modulation such as Quadrature Amplitude Modulation (QAM).
\appendices

\section{Generalized Bessel Functions and the CE-OFDM Waveform Model}
\label{sec:AppendixI}
The $L$-dimensional, $(L-1)$-parameter GBF possesses the following generating function \cite{DattoliII}
\begin{multline}
\exp\left[\frac{z_1}{2}\left(q - \frac{1}{q}\right) + \sum_{\ell=2}^L\frac{z_{\ell}}{2}\left(q^{\ell}\gamma_{\ell-1} - \frac{1}{q^{\ell}\gamma_{\ell-1}}\right)\right] \\ = \sum_{m=-\infty}^{\infty}\mathcal{J}_m\left(\left\{z_{\ell}\right\}_{\ell=1}^L;\left\{\gamma_{\ell}\right\}_{\ell=2}^L\right)q^m
\label{eq:gbfGenFunc}
\end{multline}
where $\{z_{\ell}\}_{\ell=1}^L$ are the $L$ variables and $\{\gamma_{\ell}\}_{\ell=2}^L$ are the $(L-1)$ parameters for the GBF as shown in \eqref{eq:sGBF}.  Letting $q = je^{j\theta}e^{j\phi_1}$ and $\gamma_{\ell} = -j^{-\left(\ell-1\right)}e^{-j\ell\phi_1}e^{j\phi_{\ell}}$ results in 
\begin{equation}
\left(q-\frac{1}{q}\right) = je^{j\theta}e^{j\phi_1}+je^{-j\theta}e^{-j\phi_1}=2j\cos\left(\theta+\phi_1\right) \\
\label{eq:q1}
\end{equation}
and
\begin{align}
\left(q^{\ell}\gamma_{\ell-1}-\frac{1}{q^{\ell}\gamma_{\ell-1}}\right) &= je^{j\ell\theta}e^{j\phi_{\ell}} + je^{-j\ell\theta}e^{-j\phi_{\ell}} \IEEEnonumber \\ &= 2j\cos\left(\ell\theta + \phi_{\ell}\right).
\label{eq:q2}
\end{align}
Inserting the results of \eqref{eq:q1} and \eqref{eq:q2} into \eqref{eq:gbfGenFunc} results in the Jacobi-Anger expansion for the $L$-dimensional $(L-1)$-parameter GBF \cite{DattoliII}
\begin{multline}
\exp\left[j\sum_{\ell=1}^L z_{\ell}\cos\left(\ell\theta + \phi_{\ell} \right) \right] = \\ \sum_{m=-\infty}^{\infty}j^me^{jm\phi_1}\mathcal{J}_m\left(\left\{z_{\ell}\right\}_{\ell=1}^L;\left\{\gamma_{\ell}\right\}_{\ell=2}^L\right)e^{jm\theta}.
\end{multline}

\section{Derivation of the CE-OFDM Waveform's Spectrum and AF}
\label{sec:AppendixII}
\subsection{CE-OFDM Spectrum}
\label{sec:AII_1}
Using \eqref{eq:sGBF}, the Fourier transform of the CE-OFDM waveform is expressed as 
\begin{equation}
S\left(f\right) = \dfrac{1}{\sqrt{T}}\sum_{m=-\infty}^{\infty}j^me^{jm\phi_1} \mathcal{J}_m \int_{-T/2}^{T/2}e^{-j2\pi\left(f-\frac{m}{T}\right)t}dt.
\label{eq:spec1}
\end{equation}
The integral in \eqref{eq:spec1} evaluates to $T\sinc\left[\pi T\left(f-m/T\right)\right]$ resulting in the final expression
\begin{equation}
S\left(f\right) = \sqrt{T}\sum_{m=-\infty}^{\infty}j^me^{jm\phi_1} \mathcal{J}_m \sinc\left[\pi T\left(f-\frac{m}{T}\right)\right].
\end{equation}

\subsection{CE-OFDM AF}
\label{sec:AII_1}
Inserting \eqref{eq:sGBF} into \eqref{eq:AF} results in the expression
\begin{multline}
\chi\left(\tau, \nu\right) = \dfrac{1}{T}\sum_{m,n}j^{(m-n)}e^{j\phi_1(m-n)} \mathcal{J}_m \mathcal{J}_n^* e^{j\frac{\pi\left(m+n\right)\tau}{T}} \\ \times \int_{-\infty}^{\infty}\rect\left(\dfrac{t-\tau/2}{T}\right) \rect\left(\dfrac{t+\tau/2}{T}\right)e^{j2\pi\left[\nu + \left(m-n\right)\right]t} dt.
\label{eq:AF_1}
\end{multline}
The integral in \eqref{eq:AF_1} simplifies to 
\begin{equation}
\int_{-\left(\frac{T-|\tau|}{2}\right)}^{\left(\frac{T-|\tau|}{2}\right)}e^{j2\pi\left[\nu + \left(m-n\right)\right]t} dt.
\label{eq:AF_2}
\end{equation}
Evaluating \eqref{eq:AF_2} results in the final expression for the CE-OFDM waveform's AF
\begin{multline}
\chi\left(\tau,\nu\right)=\left(\frac{T-|\tau|}{T}\right) \sum_{m,n} j^{(m-n)}e^{j\phi_1(m-n)} \mathcal{J}_m \mathcal{J}_n^*
 \\ \times e^{j\frac{\pi\left(m+n\right)\tau}{T}} \sinc\left[\pi \left(\frac{T-|\tau|}{T}\right)\left(\nu T +\left(m-n\right)\right) \right].
 \label{eq:AF_3}
\end{multline}
The ACF is then found by setting $\nu=0$. 

\section{Derivation of the CE-OFDM Waveform's EOA Parameters}
\label{sec:AppendixIII} 
This section calculates the EOA parameters of the rectangularly windowed CE-OFDM waveform model in \eqref{eq:CE-OFDM}

\subsection{RMS Pulse Length}
\label{subsec:App3_1}
Inserting \eqref{eq:complexExpo} into the RMS pulse-length expression \eqref{eq:Trms} yields the expression
\begin{IEEEeqnarray}{rCl}
\tau_{rms}^2=\dfrac{4\pi^2}{T}\int_{-T/2}^{T/2}t^2dt = \dfrac{\pi^2T^2}{3}
\label{eq:tau_2}
\end{IEEEeqnarray}
which is the standard result for any rectangularly windowed FM waveform \cite{Cook}.  

\subsection{RMS Bandwidth}
\label{subsec:rms1}
The RMS bandwidth $\beta_{rms}^2$ is calculated via \eqref{eq:Brms}.  Due to the odd-symmetry of the frequency modulation function \eqref{eq:freq}, the second integral in \eqref{eq:Brms} simplifies to 0.  Inserting \eqref{eq:ceofdm_2} into \eqref{eq:Brms} results in the expression
\begin{multline}
 \beta_{rms}^2  = \dfrac{16 \pi^4 h^2}{T^3}\int_{-T/2}^{T/2} \left[\sum_{\ell=1}^L \left(-\ell\tilde{\alpha}_{\ell}\right)\sin\left(\dfrac{2\pi \ell t}{T} \right) + \right. \\ \left. \left(-\ell\tilde{\beta}_{\ell}\right)\cos\left(\dfrac{2\pi \ell t}{T} \right)  \right]^2dt.
\label{eq:brms_2}
\end{multline}
Equation \eqref{eq:brms_2} can be rewritten as
\begin{multline}
 \beta_{rms}^2 = \dfrac{8 \pi^4 h^2}{T^2} \left(\dfrac{2}{T}\right)\int_{-T/2}^{T/2} \left[\sum_{\ell=1}^L \left(-\ell\tilde{\alpha}_{\ell}\right)\sin\left(\dfrac{2\pi \ell t}{T} \right) + \right. \\ \left. \left(-\ell\tilde{\beta}_{\ell}\right)\cos\left(\dfrac{2\pi \ell t}{T} \right)  \right]^2dt.
\label{eq:brms_3}
\end{multline}
The scaling factor $2/T$ times the integral in \eqref{eq:brms_3} can be simplified by utilizing the Parseval's theorem for Fourier series resulting in the expression
\begin{IEEEeqnarray}{rCl}
\sum_{\ell=1}^L {\ell}^2\left(\tilde{\alpha}_{\ell}^2 + \tilde{\beta}_{\ell}^2\right) = \sum_{\ell}{\ell}^2.
\label{eq:brms_4}
\end{IEEEeqnarray}
Using the formula for the sum of the first $L$ integers squared yields the final expression
\begin{equation}
\beta_{rms}^2 = \dfrac{4 \pi^4 h^2}{3T^2}\left(2L^3 + 3L^2 +L \right).
\label{eq:brms_5}
\end{equation}

\subsection{Range-Doppler Coupling Factor}
\label{subsec:RDCF}
From \eqref{eq:freq}, $\dot{\varphi}\left(t\right) = -\dfrac{4\pi^2 h}{T}\sum_{{\ell}=1}^L{\vert\Gamma_{\ell}\vert\sin\left(\dfrac{2\pi \ell t}{T}+\phi_{\ell}\right)}$ and the RDCF is therefore expressed as
\begin{IEEEeqnarray}{rCl}
\rho = -\dfrac{8\pi^3h}{T^2} \sum_{\ell=1}^L\ell \vert\Gamma_{\ell}\vert\int_{-T/2}^{T/2} t \sin \left(\frac{2 \pi \ell t}{T}+\phi_{\ell}\right) dt.
\label{eq:rho_2}
\end{IEEEeqnarray}
Using the trigonometric identity $\sin\left(a\pm b\right) = \sin a \cos b \pm \cos a \sin b$, \eqref{eq:rho_2} is expressed as
\begin{multline}
\rho = -\dfrac{8\pi^3h}{T^2} \sum_{\ell=1}^L\ell \vert\Gamma_{\ell}\vert \left[\cos\left(\phi_{\ell}\right) \int_{-T/2}^{T/2}t\sin\left(\dfrac{2 \pi \ell t}{T}\right) +  \right.  \\ \left. \sin\left(\phi_{\ell}\right) \int_{-T/2}^{T/2}t\cos\left(\dfrac{2 \pi \ell t}{T}\right) \right].
\label{eq:rho_3}
\end{multline}
The second integral evaluates to zero since the integrand is an odd-symmetric function evaluated over an even-symmetric interval leaving only the first integral to evaluate.  Using integration by parts, the first integral in \eqref{eq:rho_3} evaluates to 
\begin{equation}
\left(\dfrac{-T}{2\pi \ell}\right)\left[T\cos\left(\pi\ell\right) + T\cos(\left(-\pi \ell\right) \right] =\dfrac{-T^2}{2\pi \ell}\left(-1\right)^{\ell}.
\label{eq:rho_4}
\end{equation}
Inserting the result in \eqref{eq:rho_4} back into \eqref{eq:rho_3} results in the RDCF $\rho$ for the CE-OFDM
\begin{equation}
\rho = 4\pi^2 h \sum_{\ell}^L \vert\Gamma_{\ell}\vert \left(-1\right)^{\ell}\cos\left(\phi_{\ell}\right).
\label{eq:rho_5}
\end{equation}
For a CE-OFDM waveform employing PSK (i.e, $\vert \Gamma_{\ell}\vert = 1$) on $L$ carrier modulation frequencies, $\rho$ is maximized when the $\left(-1\right)^{\ell}\cos\left(\phi_{\ell}\right)$ terms are $+1$ for all $\ell$.  This is achieved when $\phi_{\ell}$ takes the form
\begin{equation}  \phi_{\ell}= \left\{
\begin{array}{ll}
      \pi, & \ell~\text{odd} \\ \\
      
      0, &  \ell~\text{even} \\
\end{array} 
\right.
\label{eq:rho_6}
\end{equation}
resulting in the expression for $\rho_{\text{max}}$
\begin{equation}
\rho_{\text{max}} = 4\pi^2h L.
\label{eq:rho_7}
\end{equation}
Using \eqref{eq:rho2}, $\tilde{\rho}_{\text{max}}$ is expressed as
\begin{align}
\tilde{\rho}_{\text{max}} &= \dfrac{\rho_{\text{max}}}{\beta_{rms} \tau_{rms}} = \dfrac{\sqrt{3}\sqrt{3}T4\pi^2hL}{\left(\pi T\right)\left(2\pi^2h\right)\sqrt{2L^3+3L^2+L}} \IEEEnonumber \\ &= \left(\dfrac{6}{\pi}\right)\dfrac{L}{\sqrt{2L^3+3L^2+L}}.
\label{eq:rho_8}
\end{align}


\end{document}